\documentclass[aps,prb,twocolumn,groupedaddress,showpacs, preprintnumbers]{revtex4-1}
\usepackage{amsmath}
\usepackage{graphicx}
\usepackage{bm}
\usepackage{amssymb}

\begin{document}

\preprint{FERMILAB-PUB-13-375-T}

\title{Tensor renormalization group study of classical $XY$ model on the square lattice}

\author{J. F. Yu$^1$, Z. Y. Xie$^1$, Y. Meurice$^2$, Yuzhi Liu$^{2,3,4}$, A. Denbleyker$^2$, Haiyuan Zou$^2$, M. P. Qin$^1$, J. Chen$^1$, and T. Xiang$^1$\\}

\affiliation{$^1$Institute of Physics, Chinese Academy of Sciences, P.O. Box 603, Beijing 100190, P. R. China}

\affiliation{$^2$Department of Physics and Astronomy, The University of Iowa, Iowa City, Iowa 52242, USA}

\affiliation{$^3$Theoretical Physics Department, Fermi National Accelerator Laboratory, Batavia, Illinois 60510, USA}

\affiliation{$^4$Department of Physics, University of Colorado, Boulder, CO 80309, USA}


\begin{abstract}

Using the tensor renormalization group method based on the higher-order singular value decomposition, we have studied the thermodynamic properties of the continuous $XY$ model on the square lattice. The temperature dependence of the free energy, the internal energy and the specific heat agree with the Monte Carlo calculations. From the field dependence of the magnetic susceptibility, we find the Kosterlitz-Thouless transition temperature to be $0.8921\pm0.0019$, consistent with the Monte Carlo as well as the high temperature series expansion results. At the transition temperature, the critical exponent $\delta$ is estimated as $14.5$, close to the analytic value by Kosterlitz.

\end{abstract}


\maketitle


The continuous $XY$ model has attracted great interest in the study of statistical and condensed matter physics\cite{Minnhagen}. In particular, great efforts have been devoted to the investigation of topological phase transition in this model in two dimensions.
At zero temperature, this system exhibits a ferromagnetic long-range order. In high temperatures, the magnetic order is melted by the thermal fluctuation and the system is in a paramagnetic phase. At low but finite temperature, Stanley and Kaplan\cite{StanleyKaplan} showed with series expansion that this model possesses a singularity with divergent magnetic susceptibility, indicating the existence of a finite temperature phase transition. This phase transition is peculiar since, as shown by Mermin and Wagner\cite{MerminWagner}, the conventional Landau type of phase transition associated with a spontaneous breaking of continuous symmetry is not allowed at finite temperature in two dimensions.

In the early 1970s, Kosterlitz and Thouless (KT)\cite{KT} explored this model using the renormalization group method. They found that the topological vortex excitations play an important role in this system. There are two kinds of topological excitations, vortices and anti-vortices.
In high temperatures, vortices and anti-vortices are short-range correlated and the system is disordered.
In low temperatures, a vortex forms a bound state with an anti-vortex.
These vortex-anti-vortex pairs begin to condense into a quasi-long-range ordered phase below a critical temperature, leading to a topological phase transition without breaking the O(2)-symmetry of the $XY$ model.
This transition is called the KT transition.
Furthermore, they found that the low temperature topological phase is critical and the correlation length diverges.

To understand the mechanism of the KT transition\cite{KT,KT1974}, extensive investigations have been done to determine the transition temperature and the critical exponents for the two-dimensional $XY$ model.\cite{Hasenbusch2, Janke, HongRuMa, Chester, Suzuki, Okabe, Kenna,Arisue,Pernici,Arisue2,Pernici2,Comi,Comi2}
Based on the Monte Carlo simulations on the square lattice up to the length $L=2048$ and $L=65536$, and on the finite size scaling technique,  Hasenbusch\cite{Hasenbusch2} and Komura\cite{Okabe} found the transition temperature to be $T_c=0.89294(8)$ and $T_c=0.89289(5)$, respectively.
Using the high temperature expansion up to the 33rd order of the inverse temperature,
Arisue\cite{Arisue} estimated that the transition temperature is $T_c=0.89286(8)$, consistent with the Monte Carlo result.

In this work, we study the thermodynamic properties of the two-dimensional $XY$ model using the tensor renormalization group method based on the higher-order singular value decomposition (abbreviated as HOTRG), which was proposed in Ref.~[\onlinecite{Xiang}]. This method can evaluate the thermal quantities in the nearly infinite lattice limit and does not have the errors inherent in extrapolations from finite size calculations.
It has already been used for studying the classical and quantum spin models with discrete physical degrees of freedom\cite{TNS7,Xiang,ChenQN,Yannick}.
For the three dimensional Ising model, the critical temperature and the critical exponents determined with this method have already reached or even exceeded the accuracy of the most accurate Monte Carlo results published\cite{Xiang}.
This is the first time the HOTRG is applied to a continuous model.
We have evaluated the temperature dependence of the internal energy and other thermodynamic quantities and determined the critical temperature from the singularity of the magnetic susceptibility.
For comparison, we have also evaluated the temperature dependence of the free energy, the internal energy and the specific heat using the Monte Carlo simulation on the $256\times 256$ lattice.

The $XY$ model is defined by the Hamiltonian
\begin{equation}
    H = - J \sum_{\left< ij \right>}\cos(\theta_{i}-\theta_{j})-h\sum_{i}\cos\theta_{i} ,
\end{equation}
where $\left< ij \right>$ denotes the summation over the nearest neighboring sites and $\theta_i$ is the angle of the spin at site $i$.
$J$ is the coupling constant between spins, which is set to 1 for simplicity.
$h$ is the applied magnetic field.

The tensor renormalization group\cite{Xiang,TNS4,ChenQN,Yannick,TNS6,TNS7} starts by expressing the partition function or the ground state wavefunction as a tensor network state, which is a product of local tensors defined on the lattice sites.
For the $XY$ model, the normal way of constructing local tensors fails because each spin has infinite degrees of freedom.\cite{TNS7}
Recently, we proposed a novel scheme\cite{Liu} to construct the tensor representation for the $XY$ or other continuous models by utilizing the character expansions\cite{StatFieldTheo}.
Here we briefly describe the key steps in this scheme and define the local tensors for the $XY$ model.

\begin{figure}[tbp]
\begin{center}
\includegraphics[width=8cm,clip,angle=0]{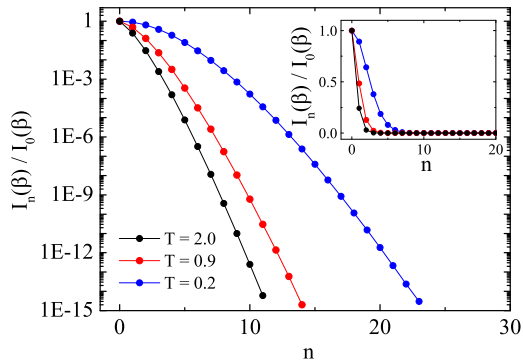}
\caption{\label{Besseli}(Color online) Variation of the modified Bessel function of the first kind $I_n(\beta)$ at three different temperatures on the logarithmic and linear (the inset) scale.}
\end{center}
\end{figure}

The partition function of the $XY$ model is given by
\begin{equation}
 Z=\int \prod_i d\theta_i e^{\beta \sum_{\left< ij \right> }\cos (\theta_i-\theta_j) + \beta h\sum_{i }\cos \theta_i },
\end{equation}
where $\beta$ is the inverse temperature.
To find its tensor representation, we take the character expansion for the Boltzmann factor\cite{StatFieldTheo}
\begin{equation}
  e^{\beta\cos{\theta}}=\sum^{\infty}_{n=-\infty}{I_n(\beta)e^{in\theta}},
\end{equation}
where $I_n(\beta)$ is the modified Bessel function of the first kind.
The partition function can then be written as
\begin{equation}\label{eqZ}
  Z=\int \prod_i d\theta_i  \prod_{n_{ij},m_i} I_{n_{ij}} (\beta) I_{m_i}(\beta h) e^{in_{ij} (\theta_i-\theta_j) + im_i\theta_i}  .
\end{equation}
By integrating out the physical degrees of freedom $\theta_i$, we can define a tensor on each lattice site
\begin{equation}\label{eqT}
T_{l,r,u,d}=\sqrt{I_l(\beta)I_r(\beta)I_u(\beta)I_d(\beta)} I_{l+u-r-d}(\beta h),
\end{equation}
where indices $(l,r,u,d)$ denote the four legs of the tensor.
The length of each leg, called bond dimension, is infinite in principle from the character expansion formula.
However, as shown in Fig. \ref{Besseli}, the series expansion coefficient $I_n(\beta)$ decreases exponentially with increasing $n$.
Thus we can truncate the series and approximate $T_{l,r,u,d}$ by a tensor with finite bond dimension $D$ with high precision.
This leads to a finite-dimensional tensor representation for the partition function
\begin{equation}
Z = \mathrm{Tr} \prod_i T_{l_i,r_i,u_i,d_i}.
\end{equation}
A bond links two local tensors. The two bond indices defined from the two end points are implicitly assumed to take the same values. For example, if the bond connecting $i$ and $j$ along the $x$ direction, then $r_i = l_j$. The trace is to sum over all bond indices.

To evaluate the partition function, we use the HOTRG to contract iteratively all local tensors\cite{Xiang}.
The HOTRG is a coarse-graining scheme of real space renormalization.
As sketched in Fig. 1 of Ref.~[\onlinecite{Xiang}], each coarse-graining step along one direction (say $y$ axis) is to contract two neighboring tensors into one, with expanded bond dimensions in the perpendicular direction ($x$ axis).
This defines a new tensor
\begin{equation}
  M^{(n)}_{l,r,u,d} = \sum_i T^{(n)}_{l_1,r_1,u,i} T^{(n)}_{l_2,r_2,i,d},
\end{equation}
where $l=l_1\otimes l_2$ and $r=r_1\otimes r_2$ are the two expanded bond indices with a
dimension $D^2$.
One can then renormalize this tensor by multiplying a unitary matrix $U^{(n)}$ on each horizontal side to truncate the expanded dimension from $D^2$ to $D$,
\begin{equation}
  T^{(n+1)}_{l^\prime,r^\prime,u,d} = \sum_{l,r}U^{(n)}_{l,l^\prime}M^{(n)}_{l,r,u,d} U^{(n)}_{r,r^\prime},
\end{equation}
where $U^{(n)}$ is determined by the higher-order singular value decomposition of the expanded tensor $M$.
The superscript $n$ denotes the $n$-th coarse-graining step.
A new tensor with a reduced bond dimension is thus obtained, and the lattice size is reduced by half.
Naturally, the truncation introduces errors, which can be reduced by increasing the bond dimension $D$.

\begin{figure}[tbp]
\begin{center}
\includegraphics[width=8cm,clip,angle=0]{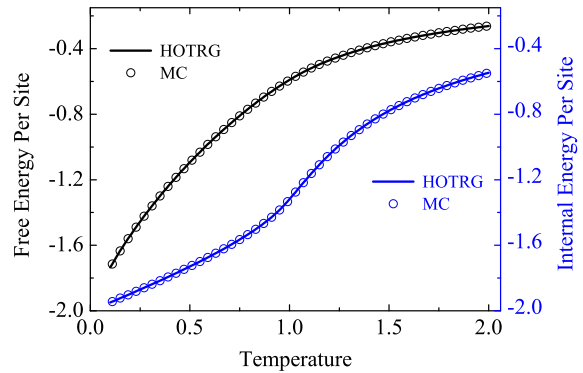}
\caption{\label{FEnergyEnergy}(Color online) Temperature dependence of the free energy (black) and the internal energy (blue) obtained by the HOTRG with $D=40$.  The free energy and the internal energy obtained by the Monte Carlo simulation on the $N = 256 \times 256$ lattice (circles) are also shown for comparison.}
\end{center}
\end{figure}

Iterating the above process along the $X$ and $Y$ directions alternately, we can finally obtain the value of the partition function and the free energy.
Considering the translation invariance, the internal energy and the magnetization can be determined by evaluating the expectation values of the local Hamiltonian and the local magnetization, respectively.
A detailed discussion on this is given in Refs.~[\onlinecite{TNS7,Xiang}].
From the derivatives of these quantities, we can further calculate the specific heat and the magnetic susceptibility.

The HOTRG can deal with very large lattice size, because each coarse-graining renormalization step reduces the lattice size by half and the number of total steps is $\log_2{N}$ with $N$ the lattice size.
In the calculation, we terminate the coarse-graining procedure after the investigated quantity has converged.
Usually this takes about 40 steps, accordingly the system size is $2^{40}$, approximately the thermodynamic limit.

\begin{figure}[tbp]
\begin{center}
\includegraphics[width=8cm,clip,angle=0]{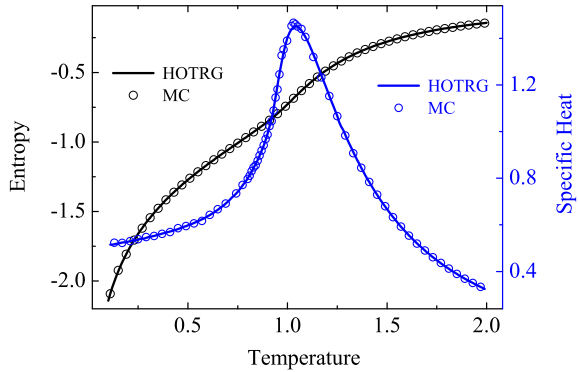}
\caption{\label{EntropySheat}(Color online) The entropy (black) and the specific heat (blue) versus temperature. Lines are from the HOTRG with $D=40$, and the circles are from the Monte Carlo simulation.}
\end{center}
\end{figure}

Figure \ref{FEnergyEnergy} shows the temperature dependence of the free energy and the internal energy obtained by the HOTRG with $D=40$ and $h=0$.
For comparison, these quantities obtained by the Monte Carlo simulation are also shown in this figure.
The HOTRG results agree with the Monte Carlo calculation.
The difference is less than $10^{-4}$ even at low temperature $T = 0.1$.
Both the internal and free energies increase smoothly from $-2$ to $0$ with increasing temperature from $0$ to infinite, and do not show any singularity at finite temperatures.
But the internal energy has more rapid increments around $T \sim 1.0$, indicating  the existence of a transition between the low and high temperature phases.

\begin{figure}[tbp]
\begin{center}
\includegraphics[width=8cm,clip,angle=0]{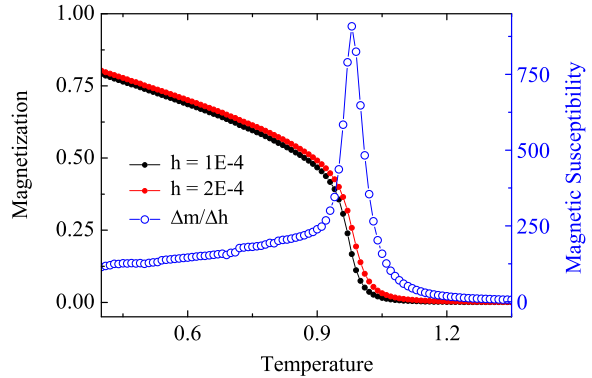}
\caption{\label{Mag}(Color online) Magnetization versus temperature with $D=40$ under two  applied magnetic fields $h=1\times10^{-4}$ (red) and $2\times10^{-4}$ (black). The magnetic susceptibility (blue) is obtained from the two magnetization curves using Eq.~(\ref{eq:chi}).}
\end{center}
\end{figure}

From the internal energy and the free energy, the temperature dependence of the entropy is deduced and depicted in Fig.~\ref{EntropySheat}.
Near temperature $T=1.0$, there is a saddle-shaped intersection.
In the same figure, the HOTRG result of the specific heat, which is obtained from the temperature derivative of the internal energy, is also shown. For comparison, the Monte Carlo result is also included, which is calculated directly from the statistical average and free from any derivative.
The results of the HOTRG and the Monte Carlo calculations agree with each other.
Only the location of the round peak is a little shifted: $T_{peak}=1.04$ for the HOTRG and $T_{peak}=1.03$ for the Monte Carlo. These peak temperatures are consistent with the Monte Carlo result($T_{peak}=1.02$) obtained by Tobochnic and Chester on the $N=30\times30$ and $60\times60$ lattices\cite{Chester}.

In our Monte Carlo calculations, the spin configurations are created with the software written by Bernd Berg\cite{bergmcmc}, and $10^6$ configurations are used for each temperature.
Because of the thermalization time, a typical saved length is $8\times10^5$.

Figure \ref{Mag} shows the temperature dependence of the magnetization in two applied magnetic fields. A continuous transition is clearly seen near $T \sim 1.0$.
At low temperature, it approaches the maximum value 1, when all spins are parallel.
With increasing temperature, the topological vortex and anti-vortex pairs are excited from the condensed phase according to the scenario of Kosterlitz and Thouless \cite{KT}.
These excitations reduce the spin-spin correlations as well as the magnetic long-range order.
The magnetization drops very quickly above $T \sim 0.9$ and approaches zero in the high temperature limit, apparently due to the thermal fluctuation.

\begin{figure}[tbp]
  \begin{center}
\includegraphics[width=8cm,clip,angle=0]{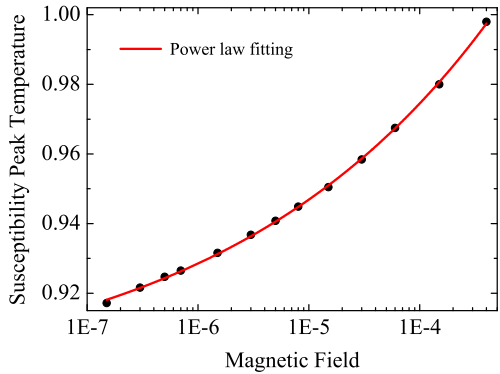}
\caption{\label{SusceptPeak} The peak temperature of the magnetic susceptibility versus the external magnetic field for $D=40$. The curve is a power fit to the peak temperature with Eq. (\ref{eqTc}). The critical temperature is extrapolated to be $0.8921\pm0.0019$.}
  \end{center}
\end{figure}

From the magnetization, the magnetic susceptibility can be evaluated using the formula:
\begin{equation}
	\chi=\frac{\Delta m}{\Delta h} = \frac{m(h_2)-m(h_1)}{h_2-h_1},
\label{eq:chi}
\end{equation}
where $m(h)$ is the magnetization in an applied field $h$. The result is shown in Fig. \ref{Mag}.
In a finite field, there is no singularity in the magnetic susceptibility and the critical divergence is replaced by a sharp peak.
When $h \sim 1.5 \times 10^{-4}$, the peak appears at $T=0.98$. Above this temperature, the magnetic susceptibility decays to zero exponentially, indicating the absence of any magnetic order in high temperatures.
The critical temperature $T_c$ can be obtained by extrapolating the peak temperature with respect to the applied field to the limit $h\rightarrow 0$.
Fig.~\ref{SusceptPeak} shows the peak temperature as a function of $h$ obtained by the HOTRG with $D=40$.
Because of the strong fluctuations near the critical point, it is difficult to determine the peak position in the extremely small field limit.
By fitting the data with the formula
\begin{equation}\label{eqTc}
  T = T_c + ah^b,
\end{equation}
we find that $T_c=0.8921\pm0.0019$, $a=0.4209\pm0.0124$ and $b=0.1768\pm0.0054$.
This value of $T_c$ agrees within the error bar with the critical temperature obtained by the Monte Carlo simulation, $T_c=0.89294(8)$, and by the high temperature expansion, $T_c = 0.89286(8)$.

\begin{figure}[tbp]
  \begin{center}
	\includegraphics[width=8cm,clip,angle=0]{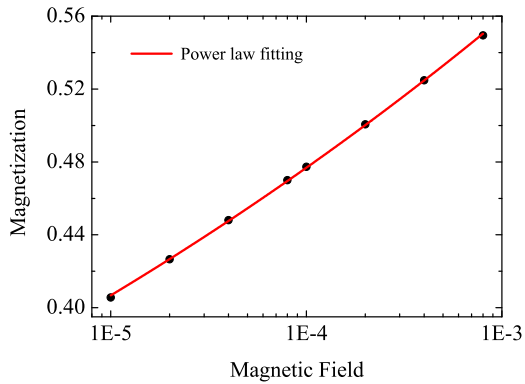}
	\caption{\label{Fitting} (Color online) Power law fit for the field dependence of the magnetization at $T_c$ with formula (\ref{eqCE}).}
  \end{center}
\end{figure}

Exactly at $T_c$, the magnetization $m$ scales with the applied field $h$ in a power law\cite{KT1974}
\begin{equation}\label{eqCE}
  m \sim h^{1/\delta}. 
\end{equation}
We have determined the value of this critical exponent $\delta$ by fitting the field $h$ dependence of the magnetization $m$ at the above estimated $T_c$.
Fig. \ref{Fitting} shows the power law fitting curve, from which we find that $\delta \approx 14.5$, consistent with the result suggested by Kosterlitz\cite{KT1974}, $\delta = 15$.

In summary, we have studied the thermodynamic properties of the continuous $XY$ model using the HOTRG on the square lattice.
From the field dependence of the magnetic susceptibility, we find that the critical temperature is about $T_c=0.8921\pm0.0019$, consistent with the Monte Carlo\cite{Hasenbusch2} and the high temperature series expansion\cite{Arisue} results.
The magnetic critical exponent $\delta$ at $T_c$ is found to be $14.5$, also in good agreement with the analytic result obtained by Kosterlitz\cite{KT1974}.

This work was supported by the Chinese Academy of Sciences Fellowship for Young International Scientists (Grant No. 2012013). Y. Meurice was supported by the Department of Energy under Award Numbers DE-SC0010114 and FG02-91ER40664. YL is supported by the URA Visiting Scholars' program. Fermilab is operated by Fermi Research Alliance, LLC, under Contract No.~DE-AC02-07CH11359 with the United States Department of Energy.

\bibliography{XYdraft}

\end{document}